\documentclass{aa}
\usepackage{txfonts}
\usepackage{epsfig}
\usepackage{subfigure}

\usepackage{times}
\usepackage{natbib}
\bibpunct{(}{)}{;}{a}{}{,}

\begin{document}

\title{XMM-Newton observation of Mrk 110}

\author{Th. Boller\inst{1}, I. Balestra\inst{1}, and W. Kollatschny\inst{2}
}

\offprints{Th. Boller, \email{bol@mpe.mpg.de}}

\institute{Max-Planck-Institut f\"ur Extraterrestrische Physik, Giessenbachstra{\ss}e,
Postfach 1312, 85741 Garching, Germany
\and Universit\"ats-Sternwarte G\"ottingen, Geismarlandstra{\ss}e 11,
37083 G\"ottingen, Germany}

\date{Received 2006, September 10/ Accepted 2006, November 26}

\authorrunning{Th. Boller, I. Balestra and W. Kollatschny}

\abstract {We report on the first XMM-Newton observation of the bright
Narrow-Line Seyfert~1 galaxy Mrk~110.}
{Our analysis is aimed to study the properties of the X-ray spectrum of Mrk~110
and compare them with those inferred from optical spectroscopy.}
{We make use of detailed timing and spectral analysis as well as high resolution
X-ray spectroscopy with the XMM-Newton gratings.}
{We find a narrow Fe~K fluorescent line,
a broad component  ($\rm FWHM\simeq16500$~km~s$^{-1}$) of the O~VII triplet,
either due to infall motions or gravitational redshift effects in the vicinity
of the central black hole,  a Comptonized accretion disc layer, and a strong starburst component.}
{We found that Mrk~110 has a complex X-ray spectrum, exhibiting
relatively strong broadening of the O~VII emission line, probably associated to X-ray emission
from the Broad Line Region (BLR), which might be correlated with the optical gravitationally redshifted,
asymmetric line profiles.
Spectral fits including a Gaussian line or a discline  give
the same statistical significance. If the broad redshifted soft X-ray components are due to gravitational
redshift effects, the distance of the line emitting regions ranges between about 0.2 and 1 light day with
respect to the central black hole.
In addition, the EPIC pn spectrum shows a double power-law and a strong starburst component. One power-law
component exhibits a photon index slope of  $\rm 1.40^{1.50}_{1.38}$, while the second is much steeper with
a power-law slope of  $\rm 2.50^{2.63}_{2.48}$. The second power-law is most probably due to thermal
Comptonization of a hot electron layer above the accretion disc.
Mrk 110 is another example of extragalactic
sources showing Comptonization effects in the accretion disc and its properties are very similar to
the Narrow-Line Seyfert 1 Galaxy Ton~S~180.
}


\maketitle

\section{Introduction}

Narrow Line Seyfert 1 galaxies, identified by the unusual narrowness of their H$\beta$ lines, are believed to be powered by
super-massive black holes (SMBH) of relatively small masses, with high accretion rates,
possibly close to the Eddington limit (e.g. Pound et al. 1995; Boller et al. 1996, Tanaka, Boller and Gallo
2005).
Furthermore, NLS1 have long been known to be characterized by extreme properties
of their X-ray emission: a strong soft excess in the {\em ROSAT} soft band
($0.1-2.4$~keV; Boller et al. 1996), unusually steep X-ray spectra in the hard
X-ray band ($2-10$~keV, Brandt 1997; Vaughan et al. 1999), very rapid and large
variability (Leighly 1999, Boller 2002 et al.).

Recent XMM-Newton detailed spectral studies have revealed more unusual spectral
properties, most notably in the form of sharp spectral drops above 7~keV, the most
extreme examples are found in 1H~$0707-495$ (Boller et al. 2002) and in IRAS~13224--3809 (Boller et al. 2003a).
 These features are sharp (within the EPIC pn resolution of about 200 eV) and time-variable
(Gallo et al. 2004). Detailed spectral modeling of these data so far have suggested
that either partial covering (Boller et al. 2002, Tanaka et al. 2004; Gallo et al. 2004) or ionized
reflection-dominated discs with light bending effects (Fabian et al. 2004;
Miniutti et al. 2003) can explain the observed features in these objects as well as in other
bright Seyfert galaxies. Within the partial covering model, variability is induced by rapid
changes in the covering fraction of the absorbers. In the light bending model, instead,
variability is essentially produced by a change of the distance
of a compact source, emitting a power-law spectrum, to the central
black hole.

Mrk~110 is a low-redshift ($\rm z=0.03529$) X-ray bright
NLS1s.
Studies of the kinematics in the central Broad Line Region (BLR) of Mrk~110
based on the variability of the Balmer and Helium (He~I, He~II) emission lines,
indicated a connection between the BLR and the central accretion disc
as well as gravitational redshift effects (Kollatschny 2004).
From the detection of gravitational redshifted emission in the variable fraction
of all broad optical emission lines in Mrk~110, a central black hole mass of
$\rm M_{grav} = (14 \pm  3)\ \cdot 10^7 M_{sun}$ is obtained.

Throughout this paper we use the following cosmological parameters:
$H_0=70$ km s$^{-1}$ Mpc$^{-1}$, $\Lambda_0=0.7$ and $q_0=0$. Error estimations for the XMM-Newton
data analysis are given for the 90 per cent confidence range.

\section{Observation and Data analysis}

Mrk~110 was observed with XMM-Newton
slightly off-axis ($\sim2'$) for $\sim48$~ks on 2004 November 15 (revolution 0904).
During this time the
EPIC pn (Str\"uder et al. 2001) and the MOS1 and MOS2 (Turner et al. 2001)
cameras, as well as the Optical Monitor OM (Mason et al. 2001) and the Reflection
Grating Spectrometers (RGS1 and RGS2) collected data.
The EPIC-PN and MOS1 cameras were operated in small window mode to reduce the pile-up,
given the brightness of this source (a few counts $\rm s^{-1}$), and utilized the thin filter.
The MOS2 camera was operated in full-frame mode, utilizing the medium filter.

The Observation Data Files were processed to produce calibrated event lists
using the XMM-Newton Science Analysis System ({\tt SAS v6.5.0}).
Unwanted hot, dead, or flickering pixels were removed as were events due to
electronic noise. Event energies were corrected for charge-transfer losses,
and EPIC response matrices were generated using the {\tt SAS} tasks
{\tt ARFGEN} and {\tt RMFGEN}. Light curves were extracted from these event
lists to search for periods of high background flaring. We cut time intervals
with count rates higher than 1.0 and 0.35 counts~s$^{-1}$ for the PN and
MOS1, respectively. The total good exposure times selected for the PN and MOS1
were $\sim33$ and $\sim46$~ks, respectively.

The source photons were extracted from a circular region with a radius of
$64''$, centered at the nominal source position. Since the PN
and MOS1 were operated in the small window mode,
we made use of the blank field
background files taken from the
XMM-SOC\footnote{ftp://xmm.vilspa.esa.es/pub/ccf/constituents/extras/background/}.
Single and double events were selected for the PN detector, and single-quadruple
events were selected for the MOS. The resulting PHA files were grouped with a
minimum of 20 counts per bin.
Pile-up effects were determined to be negligible for the PN, but not for the MOS,
given the brightness of the source ($\sim5$ counts~s$^{-1}$, the threshold
for a 1\% pile-up being 0.70 counts $\rm s^{-1}$ in the full-frame mode and 5.0 counts $\rm s^{-1}$ in the
small-window mode). Therefore the MOS data were excluded
from the analysis presented in this paper.

The RGS were operated in standard Spectro+Q mode. The first- and second-order
RGS spectra were extracted, both for RGS1 and RGS2, using the {\tt SAS} task
{\tt rgsproc}, and the response matrices were generated using {\tt rgsrmfgen}.

The OM was operated in imaging mode for the entire observation.
images were taken in three filters:  in $UVW1$ ($245-320$ nm),  in
$UVM2$ ($205-245$ nm), and  in $UVW2$ ($180-225$ nm).
We will not discuss the OM data in this paper.

\section{Timing properties}

During the observation the source flux varied by less than 15\% both in the soft
($0.3-2$~keV) and hard ($2-10$~keV) band, with no significant spectral variations,
as shown from the hardness ratio (see Fig. 1). Therefore our spectral
analysis is performed on the spectrum integrated over the whole good exposure time.

\begin{figure}
\centering
\includegraphics[width=6.5 cm, angle=270]{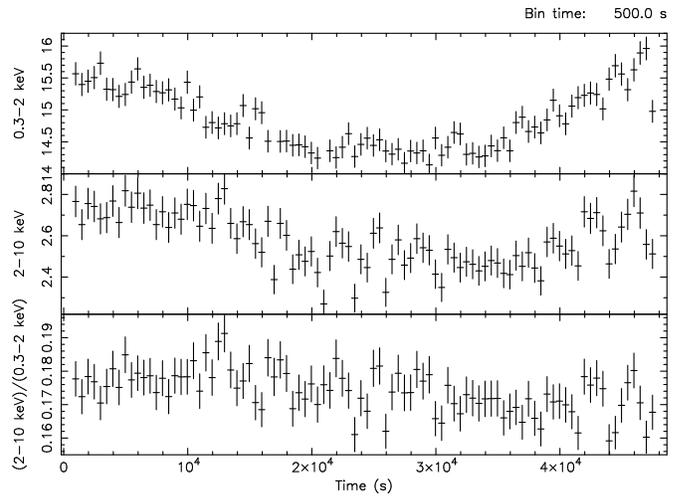}
\caption{EPIC-pn light curves in the $0.3-2$~keV (upper panel)
and $2-10$~keV band (middle panel) together with the hardness ratio
($2-10$~keV$/0.3-2$~keV) as a function of time (lower panel).}
\label{hratio}
\end{figure}

\section{Broad band spectral complexity}

The EPIC-pn  spectrum of Mrk~110 has a relatively high
signal-to-noise ratio (S/N) up to high energies.
The spectrum of the source is clearly detected above the background up to
$\rm 11$~keV (at less than 11~keV the source count rate is still
$3\sigma$ above the background as shown in Fig. 2).

\begin{figure}
 \includegraphics[width=6.5 cm, angle=270]{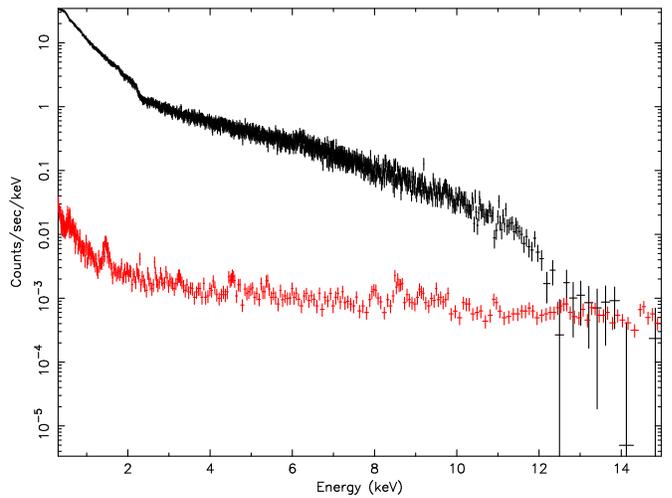}
 \caption{EPIC pn source and background spectrum of Mrk 110. The source is well
 above the 3 $\rm \sigma$ level up to 11 keV.
 }
\end{figure}

\begin{figure}
 \includegraphics[width=6.5 cm, angle=270]{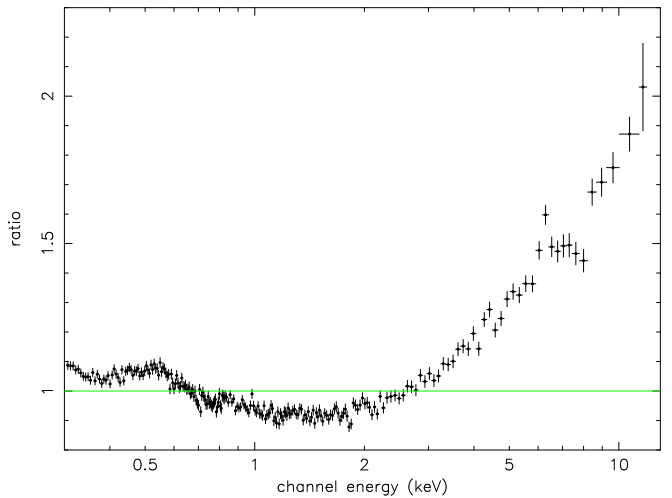}
 \caption{Ratio data-to-model for a simple power-law fit to Mrk 110. The fit is statistically not acceptable
 and reveals strong spectral complexity.
 }
\end{figure}

If the EPIC-pn broad band ($0.3-11$~keV) spectrum was first fitted with a simple power law.
The result is a statistically very poor fit ($\chi^2=7579.9$ for 1732 d.o.f..).
This is indicative for strong spectral complexity. Strong soft X-ray
excess emission is usually expected as well as steep hard photon indices (sometimes
sharp spectral drops above 7 keV, (Boller et al. 2002, 2003a, Gallo et al. 2004,
Tanaka, Boller, Gallo 2005)).
In the following we examine these spectral parameters to provide
an acceptable spectral fit and to reveal the underlying physical processes.

\subsection{The baseline model: Comptonized accretion disc  and collisionally ionized plasma emission
}


\begin{figure}
 \includegraphics[width=6.5 cm,height=8.8cm,  angle=270]{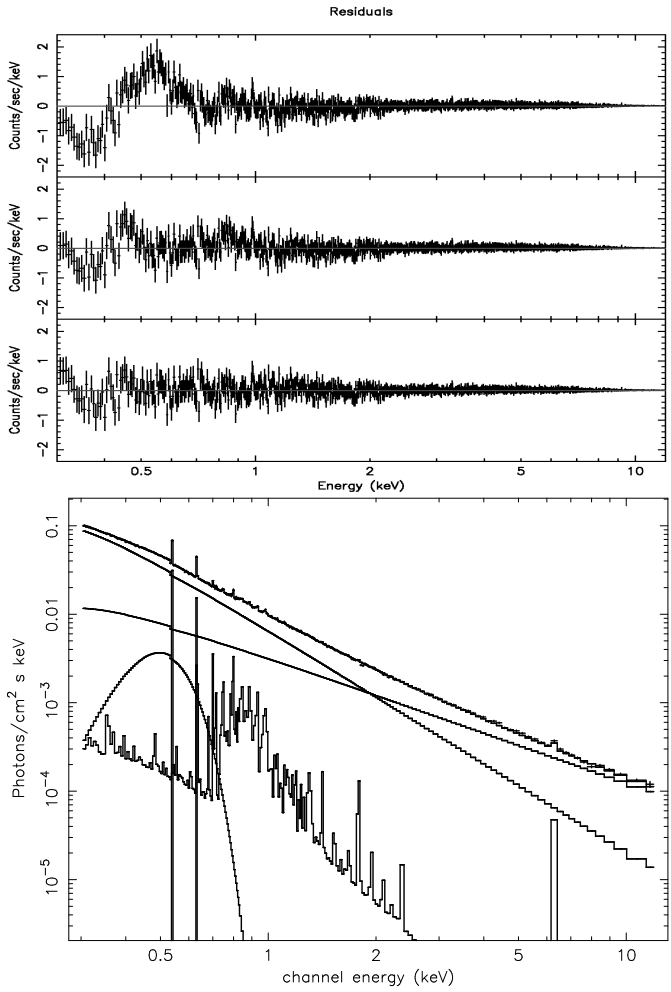}
  \includegraphics[width=6.5 cm, height=8.8cm, angle=270]{two_pl_oxy_mekal_cal_uf_2808.ps}
 \caption{{\bf Upper:} Residua for a double power-law fit (upper panel), a double power-law plus
 two broad and redshifted O~VII and O~VIII lines (middle panel), and  residua for the best fitting model
 obtained by adding a Gaussian
 line in the energy range between 0.4 and 0.5 keV to account for calibration uncertainties and adding
 a collisionally ionized plasma emission (lower panel).
 {\bf Lower:} Unfolded model for the best fitting spectrum. The corresponding residua are shown in the
 upper lower panel (see text for details).
 }
\end{figure}

We have fitted the EPIC pn data with two power-law
models and a narrow unresolved Fe K line.
The foreground absorption value is always fixed to the Galactic value of
$\rm N_H= 1.47 \cdot   10^{20}\ cm^{-2}$.
The residua are shown in the upper panel
of Fig. 4. While the energy band above 1 keV is well fitted
with these components, strong deviations are apparent between 0.3 and 1.0 keV. The $\rm \chi^2$ value is
2151.3 for 1730 d.o.f., resulting in a reduced $\rm \chi^2_{red}$ value of 1.24. A strong excess is found
between about 0.45 and 0.7 keV in the observers frame as well as a significant underestimation of the
data below 0.45 keV. From the detailed analysis of the RGS grating data,  presented in Sect.~5,
the presence of a significant, broad and redshifted emission line, most probably associated with
the O~VII triplet, has been revealed (the significance of the broad line is above the $\rm 3\ \sigma$ limit).
Also an indication for the presence of a broad O~VIII line
is found (between the 2 and 3 $\rm \sigma$ limit).
In order to take into account the presence of the broad O~VII and O~VIII lines detected with the
RGS we have added two
Gaussian lines with the energies as detected with the RGS gratings and fixing  the line widths in the fit.
This significantly improves the fit (the $\rm \chi^2$ value is 1858.6 for 1728 d.o.f., resulting
into a reduced $\rm \chi^2_{red}$ value of 1.08; upper middle panel of Fig. 4). Between 0.5 keV and 0.8 keV the model is now consistent
with the data. However, strong deviations remain below 0.5 keV and between about 0.7 and 1 keV
(c.f. Fig. 4 upper middle panel).
To account for these residua we have added two model components.
Firstly, for observations
in the small window mode it is known  that calibration uncertainties occur between 0.4 and 0.5~keV
(XMM-SOC-CAL-TN\_0052 (2006)). The deviations reach 10\% of the flux in this energy range.
To account for this calibration uncertainties we have added a Gaussian line and leaving the energy
free in the band between 0.4 and 0.5 keV. The corresponding line width is also a free parameter.
Secondly, we have added a collisionally ionized plasma emission to the model
(strong starburst emission is expected in NLS1s; Mathur 2000).
The $\rm \Delta \chi^2$ value is
44 for 2 d.o.f. This improves the fit above the 3$\rm \sigma$ limit.
The presence of the starburst component is further supported by the detection of an excess emission between
about 0.7 and 1.0 keV in the EPIC pn count rate spectrum and the identification of  emissions lines
 probably associated with Ne~IX, Fe~XVII and O~VIII (Ly-$\rm \beta$ and Ly-$\rm \delta$) between 0.7 and 1.0 keV in the observers frame (see also
 Fig. 5).
The best fitting model we  obtained has a reduced $\rm \chi^2_{red}$ value of 1.02 for 1723 d.o.f. (Fig. 4, lower panel).
The value for the photon indices of the power-law components are $\rm 1.40^{1.50}_{1.38}$ and
$\rm 2.5^{2.63}_{2.48}$, respectively. The flat power-law component is quite consistent with the
canonical slope found in broad-line Seyfert 1 Galaxies and is assumed to arise via inverse Compton scattering
of accretion disc photons in the hot accretion disc corona. The steeper power-law component is
most probably due to inverse Compton scattering of accretion disc photons on a hot electron layer
above the accretion disc (see below for further details).
The temperature of the plasma emission is $\rm 0.71^{0.81}_{0.63}$ keV.
This is consistent with the lowest temperature component found in ULRIGs (e.g. NGC 6240, Boller et al. 2003)
and in starburst galaxies (e.g. NGC 253; Pietsch et al. 2001).
The abundance remains unconstrained and was therefore fixed to solar metallicities.

To achieve a more physical understanding of the origin of the steep power-law photon index, we have
replaced the simple power-law component with the compTT model in XSPEC (the soft excess is modeled by
Comptonization of a thermal spectrum from the accretion disc).
The statistical significance of the two models is the same.
The temperature of the thermal spectrum is $\rm kT_{bb} = 34^{39}_{11}$ eV.
The values for the temperature of the hot electron layer and the optical depth are
kT = $\rm 59^{80}_{54}$ keV and $\rm \tau = 0.23^{0.28}_{0.20}$.
The resulting $\rm \gamma$ value is 0.9 for a photon index of 2.5.
To our knowledge no other physical mechanisms, apart from disc comptonization,
are able to produce the steep power-law spectral component.
Miniutti, Fabian and Goyder et al. (2003) have proposed the presence of a compact source
above the accretion disc (the reflection model) emitting a power-law continuum. However, the
physical interpretation of the power-law component is not discussed.

A narrow unresolved Fe K line is detected at an energy of $\rm 6.40^{6.42}_{6.36} keV$
with an equivalent width of $\rm 52^{70}_{36}$ eV.
The line width remains unresolved within the energy resolution of the EPIC pn detector.

The absorption corrected flux value in the 0.3 to 11 keV band is $\rm 6.5 \cdot 10^{-11}\ erg\ cm^{-2}\ s^{-1}$.
The corresponding luminosity is $\rm 1.8 \cdot 10^{44}\ erg\ s^{-1}$.

Disc comptonization have been discussed for other sources. The NLS1 Ton~S~180 can also be modeled
with two power-law components (Vaughan, Boller and Fabian et al. 2002). The electron temperatures
are quite similar to that of Mrk~110. Other sources are Mrk 335 (Gondoin, Orr, Lumb et al. 2002) or
PKS~0558--504 (Brinkmann, Arevalo, Gliozzi et al. 2004).

\subsection{Testing other models}

As almost all NLS1s do show multicolour disc emission (MCD emission), we have modeled the
spectral energy distribution of Mrk 110 with a black body component, an underlying power-law and
a narrow unresolved Fe K line.
Although the black body temperature-, power-law-, and Fe K line parameters are constrained,
the fit is statistically not acceptable ($\chi^2=2466$ for 1730 d.o.f.), resulting into
a reduced $\rm \chi^2_{red}$ value of 1.42.  Strong residua are apparent at energies greater than 5 keV.
We conclude, that a simple thermal emission from the accretion disc is not consistent with the spectral
energy distribution between 0.3 and 11 keV.
When replacing the black body component with an optically thin emission component (using
the apec model in XSPEC for a collisionally ionized plasma), the spectrum can also not be fitted
with the required statistical significance ($\chi^2=2466$ for 1730 d.o.f.).
Also a combined black body plus apec model does not
give an adequate spectral model to the EPIC pn spectrum of Mrk 110 ($\chi^2=2469$ for 1728 d.o.f.).

Ionized reflection may also account for the soft X-ray excess. We have modeled the EPIC pn spectrum of Mrk 110
with ionized reflection code {\tt reflion} within XSPEC. The best fitting model gives a rather
poor fit to the data ($\rm \chi^2 = 4935\ for\ 2341\ d.o.f.$).
In addition, the ionization parameter appears to be unphysically high ($\rm \xi \simeq 2030$).
We have tried a second ionizing reflection
component and found that it is not required by the EPIC pn data. Even freezing the model parameters from the single
reflection component
does not result in a better fit as the model parameters of the second component remain unconstrained.  We
therefore conclude, that ionized reflection from the disc is most probably not the underlying physical model for the
XMM-Newton X-ray spectrum of Mrk 110.

Finally we have tested whether a combination of thermal and nonthermal Comptonization of an
accretion disc layer can account for the observations using the {\tt eqpair} code from Copi (1992).
The model describes a single plasma emission with a hybrid thermal, non-thermal electron distribution
and was successfully applied to the NLS1 Ton~S~180 (Vaughan et al. 2002).
We tried to apply this model to the data as a successful fit would give important information on
the non-thermal electron distribution, e.g. the soft compactness parameter, the non-thermal compactness,
or the non-thermal injection of relativistic electrons.
Applying the hybrid comptonized plasma layer to the Mrk 110 data, again no statistically acceptable fit could
be obtained.

\section{Relativistically broadened optical and X-ray lines?}

\subsection{Spectral analysis of the RGS}

\begin{table*}
\caption{Strongest emission lines detected in the RGS spectra and their respective
broadened components. Results refer to the fit of the spectra rebinned to have at least
20 counts/bin and applying the $\chi^2-$statistics. Line energies are in the source
rest frame and line fluxes are in units of $10^{-5}$~photon~cm$^{-2}$~s$^{-1}$.
$^*$ denotes fixed parameters.}
\label{rgs_fit_chi}
\centering
\begin{tabular}{l l l l l l l l}
\hline\hline
Line               & E$_{\mathrm{lab}}$~(eV)  & E~(eV)                & $\sigma$~(eV)        & Flux                & EW~(eV) &  $\Delta\chi^2$ & $P$ \\
\hline
O VIII Ly-$\alpha$ & $652.38$                 & $653.0\pm0.3$         & $0^*$                & $3.3_{-1.8}^{+1.6}$ & $0.94_{-0.67}^{+0.63}$ &  24.2 & 0.999 \\
O VIII broad       &                          & $646.6\pm2.0$         & $<20$                & $2.4_{-1.7}^{+1.6}$ & $1.34_{-0.73}^{+0.65}$ &  6.8 & 0.985 \\
\hline
O VII 1s-2p (r)    & $572.92$                 & $572.3\pm0.8$         & $0^*$                & $3.6_{-1.7}^{+2.7}$ & $0.9_{-0.4}^{+0.7}$    &  9.5  & 0.992 \\
O VII 1s-2p (i)    & $567.92$                 & $568.3_{-0.7}^{+0.4}$ & $0^*$                & $5.1_{-3.5}^{+1.4}$ & $1.3_{-0.9}^{+0.4}$    &  8.0  & 0.987 \\
O VII 1s-2p (f)    & $559.93$                 & $561.6_{-0.4}^{+0.8}$ & $0^*$                & $4.2_{-1.8}^{+2.2}$ & $1.0_{-0.4}^{+0.5}$    & 7.1  & 0.978 \\
O VII broad        &                          & $554_{-3}^{+2}$       & $13_{-3}^{+4}$       & $40\pm7$            & $12\pm2$               &  165  & 0.999 \\
\hline
N VII Ly-$\alpha$  & $499.37$                 & $500.7\pm4.0$         & $0^*$                & $1.4\pm1.2$         & $0.35\pm0.30$          &  1.7  & 0.605 \\
N VII broad        &                          & $494.5\pm3.0$         & $3.6_{-2.0}^{+3.2}$  & $4.5_{-2.0}^{+3.5}$ & $1.1_{-0.5}^{+0.9}$    & 16.6 & 0.995 \\
\hline
C VI Ly-$\alpha$   & $366.82$                 & $367.3_{-0.7}^{+1.4}$ & $0^*$                & $1.9_{-1.8}^{+1.6}$ & $0.27_{-0.25}^{+0.23}$ &  3.2  & 0.802 \\
C VI broad         &                          & $354\pm3$             & $3.9_{-1.0}^{+3.8}$  & $11\pm5$            & $1.5\pm0.7$            &  14.5 & 0.994 \\
\hline
\end{tabular}
\end{table*}

The EPIC spectrum shows a broad feature at about $0.55$~keV (redshifted with respect to the
mean energy of the O~VII triplet). To study the soft X-ray
spectra in more detail, we examined the simultaneous XMM-Newton gratings
data of Mrk~110. Fig.~5 shows the smoothed fluxed spectrum, resulting
from the sum of the two orders of RGS1 and RGS2.
We analyzed simultaneously the binned RGS1 and RGS2 data with {\tt XSPEC}.
Here we present measurements
of the strongest emission lines showing complex profiles, that are the Oxygen triplet
(O~VII recombination (r), intercombination (i) and forbidden (f) lines at
$572.9$, $567.9$ and $559.9$~eV, respectively), the Ly-$\alpha$ lines  of O~VIII
at $652.0$~eV, N~VII at $499.4$~eV and C~VI at $366.8$~eV.
These narrow lines are unresolved and neither red- or blueshifted given the
present data quality.
The most significant narrow line features are the
O~VII triplet and fluorescence O~VIII line.
The N~VII  and C~VI  line emission features
are below the the $\rm 3 \sigma$ limit.
For the narrow O~VII triplet we deduce a R factor of $\rm 0.82\pm0.64$.
According to Porquet and Duban (2000) this corresponds to densities ranging
from about $\rm 6\cdot 10^{10}\ to\ 5\cdot 10^{11}\ cm^{-3}$.
A detailed modeling of the complete RGS spectrum, including a
simultaneous treatment of absorption and emission lines, is beyond the
scope of this paper, and it will be treated elsewhere.

\begin{figure*}
\centering
\includegraphics[width=18.0cm, height=8cm]{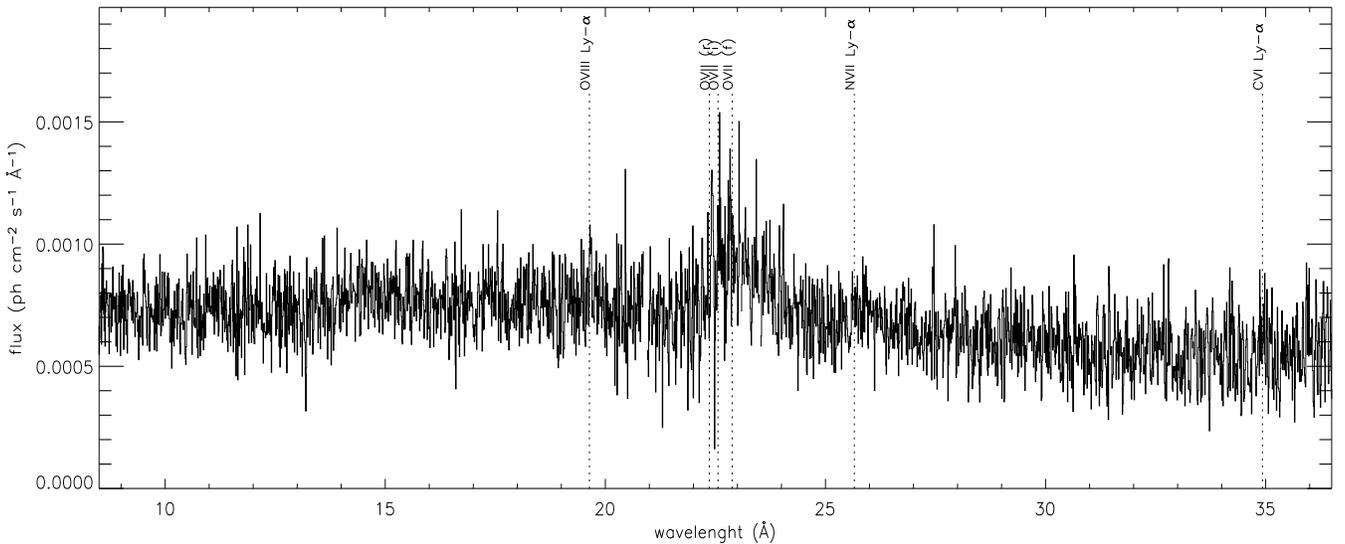}
\caption{RGS spectrum of Mrk~110. The energies at which the most
common emission lines should lie are marked with vertical dashed lines. }
\label{rgs_smooth}
\end{figure*}

\begin{figure*}
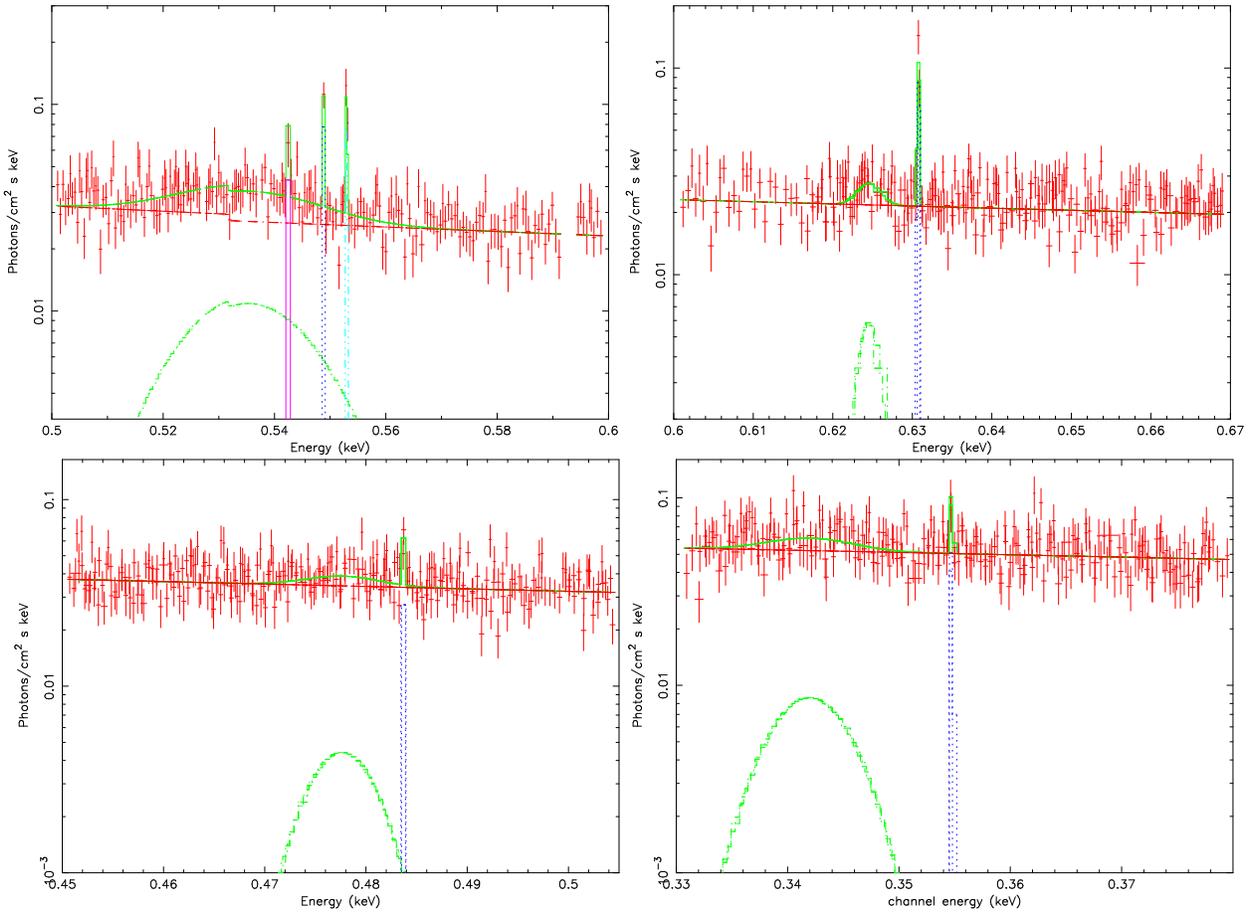

\centering
\includegraphics[width=6.0 cm, angle=-90]{OVII_bin20.ps}
\includegraphics[width=6.0 cm, angle=-90]{OVIII_bin20.ps}
\includegraphics[width=6.0 cm, angle=-90]{NVII_bin20.ps}
\includegraphics[width=6.0 cm, angle=-90]{CVI_bin20.ps}
\caption{{\bf Unfolded} RGS spectrum in the energy range of the OVII (upper left)
triplet, the O~VIII (upper right), N~VII (lower left), and C~VI (lower right panel) lines.
A relatively
strong bumpy feature is apparent around the O~VII triplet above the 3 $\rm
\sigma$ limit. Less pronounced broad
features are also detected around the Lyman-$\alpha$ of O~VIII, N~VII, and C~VI
(c.f. Table 1).
The redshift of the broad component is 0.023,
associated with the O~VII triplet.  With
the present statistic we cannot decide, whether this is due to gravitational
redshift or
to infall of matter towards the central black hole.
}
\label{rgs_OVIII}
\end{figure*}

Interestingly, we found a broad line component around the O~VII triplet (c.f. Fig. 6, left upper panel, dashed line) required by
the data (the significance is above the $\rm 3 \sigma$ limit).
The redshift is 0.023 and the FWHM of the broad component is about $\rm 16500\ km\ s^{-1}$.
We caution, however, that the broad O~VII component is related to a triplet and, therefore, gives an
overestimate of the line width due to the superposition of the three broadned O~VII lines.
We treat this value of the line width as an upper limit.
The mean rest energy of the line depends on the
relative fluxes of the triplet, which in turn depend on the
density of the emission region. Unfortunately we are unable
to constrain the triplet line ratios or the density with the
present data quality, therefore we assume a representative ratio
of the recombination (r), intercombination (i) and  forbidden (f) lines.
Assuming that the displacement of the broad line it is due to gravitational
redshift effects ( following M\"uller and  Wold (A\&A, 2006)), we would infer a distance of about 30 $\rm R_S$
of the line emitting region to the central black hole .
The presence of relativistically broadened soft X-ray lines in MCG-6-30-15 or in Mrk 766 have already been
discussed by Branduardi-Raymont et al. (2001),  Mason et al. (2003)
or Sako et al. (2003).
Ogle et al. (2004) reported on the discovery
of a relativistically broadened O~VIII line in NGC~4051.
The XMM-Newton observation of Mrk 110 shows indications for broad redshifted components also for the O~VIII, N~VII and  C~VI lines
with significance levels between 2 and 3 $\sigma$.
Presently, we cannot decide whether the redshift of the broad
component, associated with the O~VII triplet, is due to gravitational redshift effects
or to infall motions towards the central black hole.

\subsection{Comparison with optical data}

The optical spectra of Mrk~110 were studied in great detail by Kollatschny (2003a, 2003b and 2004).
Kollatschny (2003b) investigated the variable broad line components of the $\rm H\alpha$,
$ H \beta$, $ He I \lambda 5876$, and $\rm He II \lambda 4686$ lines. The shifts between the
rms profile and the mean line profiles were measured and the differential shifts $\rm \Delta v = \Delta z \cdot c$
were interpreted as gravitational redshift effects.  In addition, the cross-correlation lags $\rm \tau$ were calculated from
reverberation measurements. In Fig. 3 of Kollatschny (2003b) a relation between the redshift of the rms
profiles (in terms of velocity shifts) as a function of distance of the rms line emitting regions is found.
Using the relation of Zheng and Sulentic (1990)
($\rm M_{grav} = c^2G^{-1}R\Delta z$, where $\rm R=c\tau$)
a gravitational black hole mass of $\rm 1.4 \cdot 10^8 M_{\odot}$
is determined.
From the narrow optical lines gravitational redshift
values of 0.00025 (H$\rm \alpha$),
0.00039 (H$\rm \beta$),
0.00062 (He~I), and
0.00183 (He~II) are obtained.
As an example, the distance of the optical line emitting regions based on He~II measurements from the central
black hole originates at a distance of 3.9 light-days,
corresponding to 230 Schwarzschild radii.
To avoid confusion, we would like to remind the reader that in NLS1s the broad lines are relatively narrow, and that the
gravitational reshift values mentioned above are measured for the BLR.

Under the assumption that the broad redshifted soft X-ray lines are due to gravitational
redshift effects, we extrapolate the velocity shift into the soft X-ray regime (Fig.~7).
If longer X-ray observations will confirm the presence of relativistic effects for soft X-ray lines,
Fig. 7 could be used to test our prediction on the distance of the four soft X-ray lines to the
central black hole.

Following Kollatschny (2003b, his Fig. 2) we have plotted in Fig. 8  the velocity shift values and the FWHM for
the broad soft X-ray lines in addition to the optical measurements. With the present data quality we cannot
disentangle between gravitational redshift effects or infall motions.

\begin{figure}[t]
             \includegraphics[width=6.0cm, angle=270]{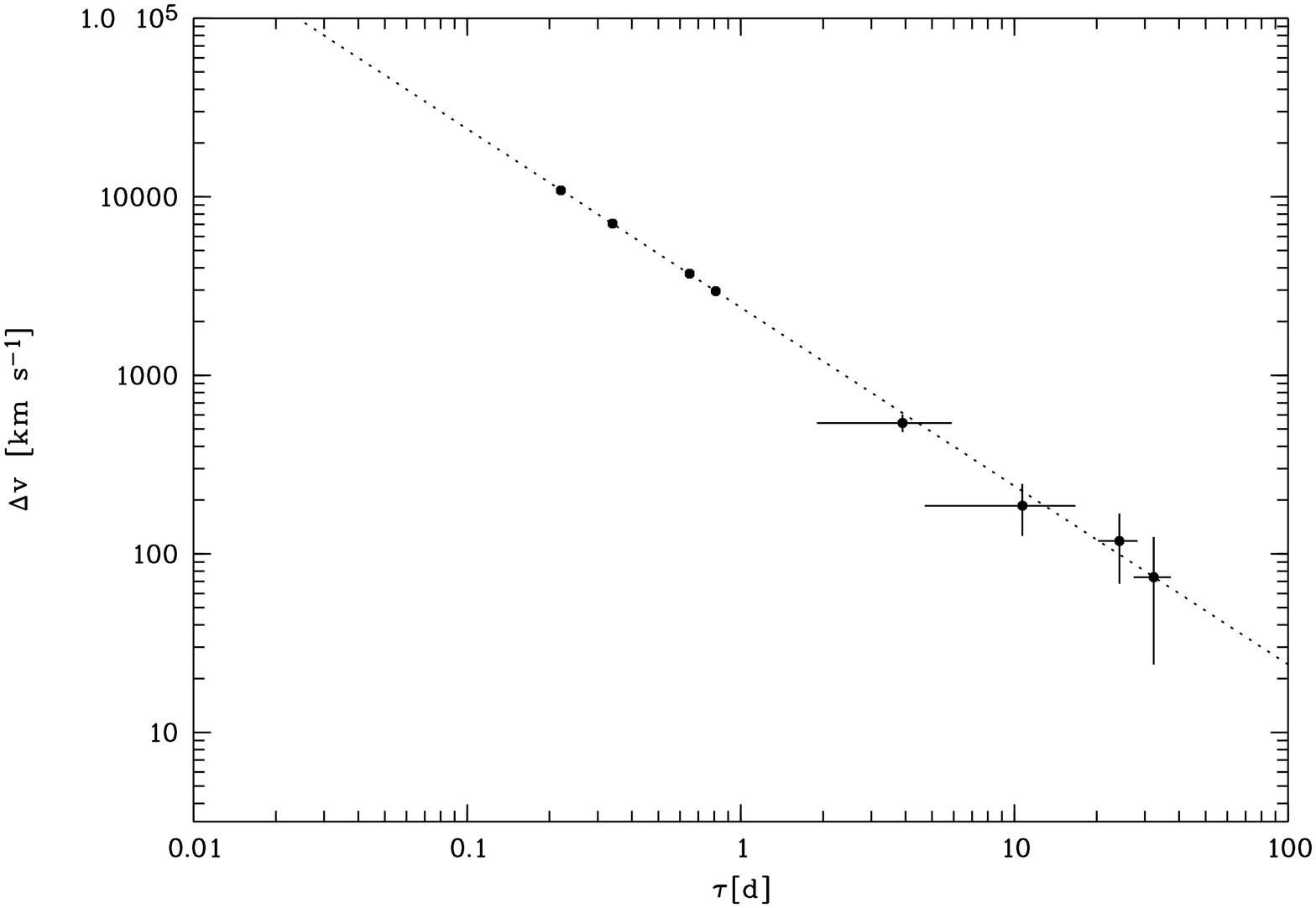}
        \caption{ Velocity shift as a function of the time lag $\tau$ in units of days ($\rm \tau = R/c$, where R is distance to the black hole).
         The points with the error bars are taken from Kollatschny (2003b). The dashed line is the Zheng and Sulentic (1990)
        relation ($\rm M_{grav} = c^2G^{-1}R\Delta z$) for a black hole mass $M_{grav} = 1.4 \cdot 10^8 M_{\odot}$. The $\rm \Delta v$ values
        are computed as $\rm c \cdot \Delta z$, where $\rm \Delta z$ is the measured redshift for the broad soft X-ray lines.
        The four dots indicate the velocity shift predicted for the  O~VIII, O~VII, N~VII, and C~VI broad line components, if the diplacement
	with respect to the corresponding narrow components is interpreted as gravitational redshift. }

\end{figure}

\begin{figure}[t]
             \includegraphics[width=6.0cm, angle=270]{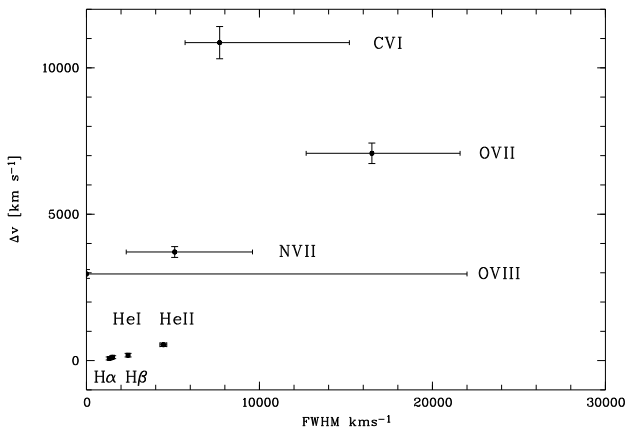}
        \caption{Velocity shift versus FWHM for the H$\alpha$, H$\beta$, He~I and He~II optical lines as discussed by Kollatschny (2003)
	 together with the soft X-ray lines (O~VIII, O~VII, N~VII, and C~VI). The velocity shifts
	 for the soft X-ray lines were calculated from the redshift ($\rm \Delta v = c\ \Delta z$)
	 from the FWHM values from the RGS broad-line profiles.}

\end{figure}

\begin{figure}[t]
             \includegraphics[width=6.0cm, angle=270]{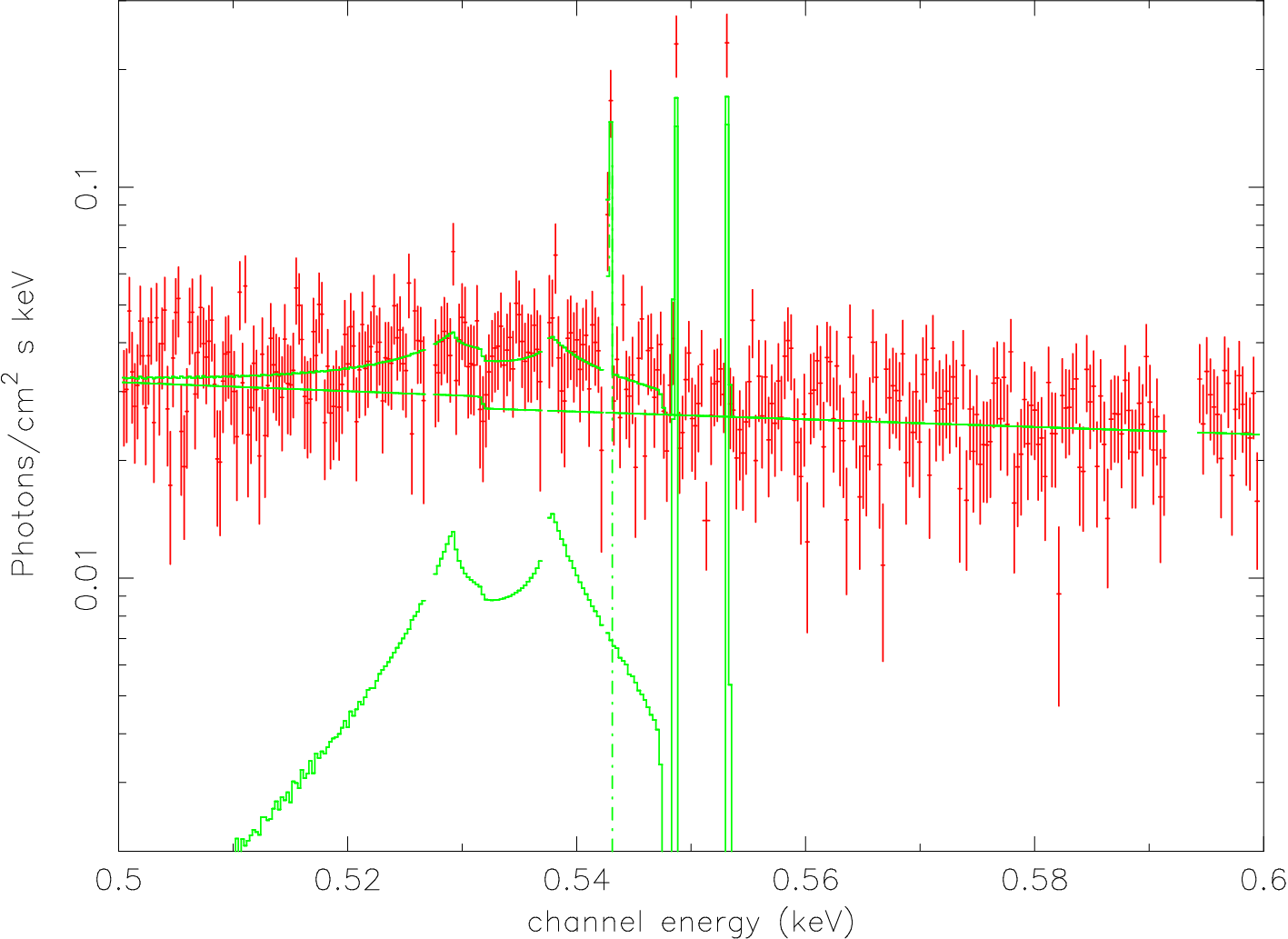}
        \caption{Unfolded spectrum for a discline fit to RGS data of the oxygen triplet. The statistical significance of the fit is the same as for
	a broad Gaussian line shown in Fig. 6. }

\end{figure}

 If the lines are optically thin one expects to see redshifted double peaked profiles due to relativistic beaming and gravitational redshift effects.
Therefore,  we have fitted the RGS data with a discline profile
for the broad redshifted O~VII component (c.f. Fig. 9). The statistical significance of the fit is equivalent to the fit with a
broad Gaussian. The best-fit parameters for the discline model are as follows:
$\rm E=531.8^{534.9}_{531.3}\ eV, \Gamma=-2.1^{-1.8}_{-2.5}, R_{in}=60^{130}_{10}\ R_G, R_{out}=2000^{5000}_{1000}\ R_G,
i=22^{39}_{18}$ degrees.  The value of $\rm \chi^2_{red}$ is 1.084 for 333 d.o.f.

Another approach is to model the broad component by fixing the restframe
    line energy and assuming the line is broadened by gravitational redshift
    and Doppler effects in a Keplerian disc.
In this way we obtain that all the
emission should come from the innermost region ($\rm 10-30\ R_G$) and the emissitity
of the disk needs to be extremly steep ($\rm -8\ to -10$) which appears unlikely.
However this could be still consistent with emission from the BLR.
The value of $\rm \chi^2_{red}$ is 1.080 for 334 d.o.f.

With the present data quality we cannot disentangle between gravitational redshift effects or just infall motions at
soft X-rays. However,
all of the putative broad components are also redshifted with respect to
the narrow emission line energies. The narrow X-ray lines from these elements
remain unresolved and the presence  of a double peaked profile remains speculative.

\section{Discussion and conclusions}

We have analyzed the XMM-Newton spectra of Mrk~110.
The EPIC pn spectrum revealed the presence of a Comptonized accretion disc
layer and the presence of a starburst component. Ionized disc models, optically
thick emission from the disc and hybrid thermal/non-thermal emission models appear
to be ruled out by the data. Mrk~110 is another active galaxy
showing a Comptonized accretion disc layer.
The optical data from Mrk~110 show asymmetric line profiles which are interpreted
as gravitational redshift effects from the central black hole.
We find  broad X-ray line emission features associated with the BLR, e.g.
the broad component of the O~VII triplet has a FWHM of $\rm \simeq16500$~km~s$^{-1}$.
The RGS data show in addition to the narrow unresolved components a broad redshifted
O~VII line.
The widths of the broad O~VII line is giving a clear indication that this line
originates from the BLR.
With the present data quality we cannot disentangle
between gravitational redshift effects for the X-ray lines or infall bulk
motion towards the central black hole.
Spectral fits to the redshifted O~VII line with a Gaussian profile or a discline
profile yield the same statistical significance.
Indications for broad redshifted X-ray lines
are also found for O~VIII, N~VII and C~VI.  The significance of these lines is below
the $\rm 3 \sigma$ limit.
Longer X-ray observations or future missions are required to
 distinguish between broad Gaussian lines or gravitationally distorted line profiles
in the soft X-ray regime.

\begin{acknowledgements}
Based on observations obtained with XMM-Newton, an ESA science
mission with instruments and contributions directly funded by ESA
Member States and the USA (NASA).
We thank the anonymous referee for very helpful comments to
improve the paper.
\end{acknowledgements}

\bibliographystyle{aa}
\bibliography{sbs}

\end{document}